\documentclass[aip,
 amsmath,amssymb,
 reprint,%
]{revtex4-1}

\usepackage{graphicx}
\usepackage{color}
\usepackage{bm}
\usepackage{amsmath}
\usepackage{amssymb}
\usepackage{hyperref}
\usepackage{float}

\def\xcm3{\mbox{cm}^{-3}}

\def\Wxcm2{\mbox{Wcm}^{-2}}

\def\Axm2{\mbox{Am}^{-2}}

\definecolor{red}{rgb}{1,0,0}
\definecolor{blue}{rgb}{0,0,1}

\begin{document}

\preprint{Discontinuity}

\title{Collisionless tangential discontinuity between pair plasma and electron-proton plasma}

\author{M.~E.~Dieckmann}\affiliation{Department of Science and Technology (ITN), Link\"oping University, 60174 Norrk\"oping, Sweden}

\date{\today}

\begin{abstract}
We study with a one-dimensional particle-in-cell (PIC) simulation the expansion of a pair cloud into a magnetized electron-proton plasma as well as the formation and subsequent propagation of a tangential discontinuity that separates both plasmas. Its propagation speed takes the value that balances the magnetic pressure of the discontinuity against the thermal pressure of the pair cloud and the ram pressure of the protons. Protons are accelerated by the discontinuity to a speed that exceeds the fast magnetosonic speed by the factor 10. A supercritical fast magnetosonic shock forms at the front of this beam. An increasing proton temperature downstream of the shock and ahead of the discontinuity leaves the latter intact. We create the discontinuity by injecting a pair cloud at a simulation boundary into a uniform electron-proton plasma, which is permeated by a perpendicular magnetic field. Collisionless tangential discontinuities in the relativistic pair jets of X-ray binaries (microquasars) are in permanent contact with the relativistic leptons of its inner cocoon and they become sources of radio synchrotron emissions.
\end{abstract}

\maketitle

Filamentation instabilities between colliding unmagnetized or magnetized pair clouds \cite{Kazimura1998a,Sironi2009,Lemoine2010,Bret2014,DieckmannMNRAS2018,Plotnikov2018} and between initially unmagnetized counterstreaming clouds of electrons and ions \cite{Kazimura1998b,Spitkovsky2008} have been studied widely with particle-in-cell (PIC) simulations. These simulations showed that a filamentation instability rapidly thermalizes the interpenetrating plasma clouds. Strong electromagnetic fields exist only in a layer that is close to the boundary that separates the inflowing upstream plasma from the thermalized one; this layer corresponds to a shock if the collision speed is high enough. See\cite{Marcowith2016} for a review of such shocks.

Mechanisms, that can enforce the separation of a fast plasma flow from an ambient plasma at rest rather than their thermalization, are  interesting in the context of relativistic astrophysical jets \cite{Marti1997,Bromberg2011}. Their plasma is dilute, which implies that binary (Coulomb) collisions between particles are rare on the time scales of interest. We call a plasma collisionless if its dynamics is determined by the electromagnetic fields generated by the collective of the particles rather than by binary collisions.

Black hole X-ray binaries can emit such jets, in which case they are called microquasars \cite{Fender2014}. Material from the companion star is attracted by the black hole and forced onto an accretion disk. Instabilities transform some of the inner disk's kinetic and magnetic energies into thermal energy heating up the disk's corona to MeV temperatures. Large clouds of electrons and positrons form (See \cite{Yuan2014} for a review and \cite{Siegert2016} for an observation of pair annihilation lines). If we assume that the temperature of this pair cloud is relativistic at its source then its initial expansion speed should be at least mildly relativistic. Open magnetic field lines that start at the inner disk allow the pair cloud to escape from the black hole's vicinity. It flows through an ambient plasma, which is initially that of the corona followed by the stellar wind of the black hole's companion star \cite{Perucho2010} and finally the interstellar medium \cite{Bordas2009}. If the pair outflow does not interact with the ambient plasma on its way losing its kinetic energy to the slow-moving ions then it can maintain its initial speed. 

Instabilities between pair plasma and electron-ion plasma can separate both. An electron-proton plasma, which was initially spatially uniform, unmagnetized and at rest, separated itself from a relativistically moving spatially localized pair cloud in the simulation in Ref. \cite{DieckmannPoP2018a}. A filamentation instability between the particles of the pair cloud and the electrons at rest was the mechanism that drove the separation of the positive charges \cite{Warwick2017,DieckmannPoP2018b}. This instability resulted in a magnetic field structure that moved relative to the protons \cite{Pelletier2019}. Protons were accelerated by the associated convective electric field. 

\begin{figure}[ht]
\includegraphics[width=\columnwidth]{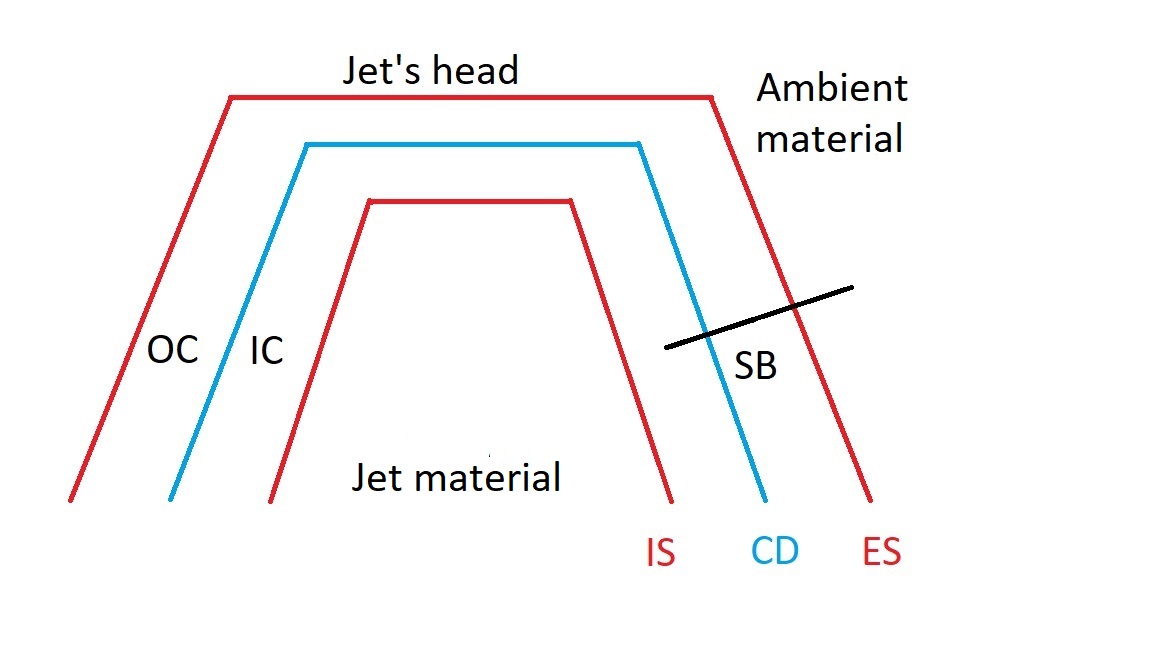}
\caption{A hydrodynamic collimated jet: The contact discontinuity (CD) separates the outer cocoon (OC) from the inner cocoon (IC). The OC contains ambient material that crossed the external shock (ES). Jets are collimated by the resistance the ambient material provides to the expansion of the ES. The CD deflects the ambient material that crossed the ES at the jet's head into the OC. Jet material, which has been shocked by its passage through the internal shock (IS), forms the IC. Its thermal pressure pushes the CD outwards. Our simulation box (SB) will be located close to the CD.}
\label{figure1}
\end{figure}

A spatially uniform magnetic field was aligned in Ref. \cite{DieckmannAA2019} with the propagation direction of a pair cloud with a limited lateral extent. This pair cloud was injected at a simulation boundary and interacted with a spatially uniform electron-proton plasma. Electromagnetic pistons emerged at the two outer boundaries of the pair cloud in the direction perpendicular to the cloud's propagation direction. Both pistons separated the positrons from the protons acting like the diagonal contact discontinuities in the sketch of a hydrodynamic jet model \cite{Marti1997,Bordas2009,Perucho2010,Bromberg2011} in Fig. \ref{figure1}.
No piston formed at the jet's head, which was mediated by a Weibel-type instability.

Once both pistons were fully developed, the pair plasma became the equivalent of the jet material in Fig.~\ref{figure1} and the electron-proton plasma the ambient one. The inner cocoon consisted of the pair plasma that was slowed down by its interaction with the piston. The high thermal pressure of the inner cocoon pushed the piston outwards creating an outer cocoon of accelerated ambient material. The costly 2D PIC simulation could, however, not be advanced to the time when internal and external shocks would form.

Here we demonstrate that we can study the formation of pistons in a one-dimensional model, which allows us to extend the simulation time way beyond that in Ref.\cite{DieckmannAA2019}. Our simulation box covers the inner cocoon, the discontinuity and the outer cocoon as indicated by the line in Fig. \ref{figure1}. A pair cloud is injected into an ambient electron-proton plasma at the boundary of the simulation box. It expands orthogonally to a magnetic field, which is spatially uniform at the simulation's start. We find electromagnetic pistons, which are tangential discontinuities in the case we consider, that grow on a time scale that is comparable to the inverse proton plasma frequency. They remain stable throughout the simulation time and separate the pair plasma from the electron-proton plasma as in Ref. \cite{DieckmannAA2019}. A supercritical fast magnetosonic shock forms at the front of the accelerated ambient protons after one inverse proton gyrofrequency. It corresponds to the external shock in Fig.~\ref{figure1}. Its downstream region will become the outer cocoon once its protons have fully thermalized. The simulation shows that the piston remains stable while the protons ahead of it heat up.

Our paper is structured as follows. Section 2 lists our initial conditions. Section 3 presents our results. Section 4 summarizes them and lists some of their astrophysical implications.   

\section{Initial and simulation conditions}

We use the EPOCH code. It solves Amp\`ere's law and Faraday's law on a numerical grid and approximates the plasma by an ensemble of computational particles (CPs). The particle currents are interpolated to the grid updating the electromagnetic fields. The latter are interpolated back to the CPs updating their velocity via the relativistic Lorentz force equation. Gauss' law and the magnetic divergence law are satisfied to round-off precision. The numerical scheme is discussed in detail in Ref. \cite{Arber2015}.

\begin{figure}
\includegraphics[width=\columnwidth]{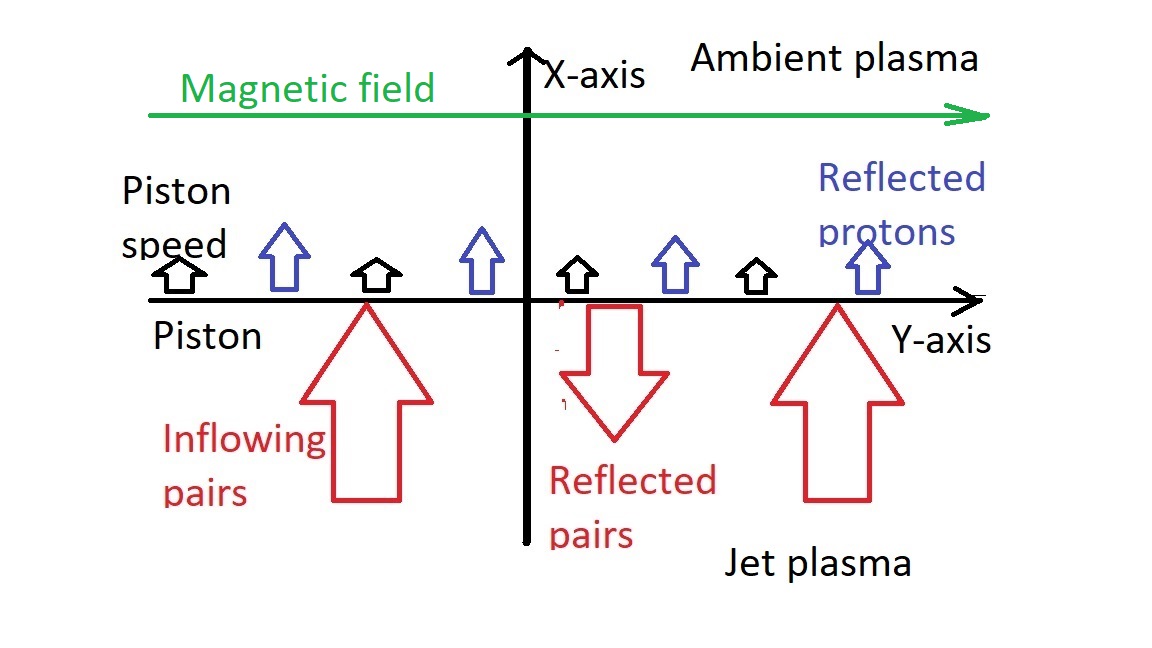}
\caption{Interval close to the electromagnetic piston, which is the central horizontal black line that coincides here with the y-axis and separates the ambient plasma from the jet plasma. Inflowing pairs are reflected by the upward moving piston and lose momentum and energy to it in the rest frame of the ambient plasma. Protons, which are reflected by the piston, propagate upwards into the ambient plasma. We align our simulation box with the vertical x-axis. A magnetic field permeates the ambient plasma and is aligned with the y-axis.}
\label{figure2}
\end{figure}

We consider here the spatial interval close to the contact discontinuity in Fig. \ref{figure1}, which becomes the electromagnetic piston in a collision-less plasma. Figure \ref{figure2} provides a close-up of this interval. Our simulation must contain a magnetized ambient plasma at rest, which consists of electrons and protons, and a pair plasma that streams towards it. We assume that the pair plasma flows along the piston normal and perpendicular to the magnetic field of the ambient plasma.

We use periodic boundary conditions and fill the box with a spatially uniform ambient plasma, which consists of electrons with the mass $m_e$ and the number density $n_0$. Protons with the same number density and the mass $m_p=1836m_e$ compensate the electron charge. Both species are initially at rest and have the temperature $T_0$ = 2 keV. The electron thermal pressure is $P_0 = n_0 k_B T_0$ ($k_B$: Boltzmann constant).

If the plasma is collisionless, the value of $n_0$ affects only the spatio-temporal scales over which the plasma processes develop but not their qualitative properties. Time is thus normalized to the electron plasma frequency $\omega_{pe}={(e^2 n_0/m_e \epsilon_0)}^{1/2}$ ($e, \epsilon_0, \mu_0$: elementary charge, vacuum permittivity and permeability) and space to the electron skin depth $\lambda_e = c/\omega_{pe}$ ($c$: speed of light). We normalize the electric field $\mathbf{E}$ to $m_e c \omega_{pe} / e$, the magnetic field to $\mathbf{B}$ to $m_e \omega_{pe} / e$ and the current density $\mathbf{J}$ to $en_0c$. We state the normalization of velocities $\mathbf{v}$ and momenta $\mathbf{p}$ explicitely in the text and figures. A magnetic field $\mathbf{B}_0 = (0,B_0,0)$ with the electron gyro-frequency $\omega_{ce}\approx0.09$ ($\omega_{ce}=eB_0/m_e\omega_{pe}$) permeates the plasma at the simulation's start $t=0$. Its normalized magnetic pressure is $P_b = B_0^2/2\mu_0P_0=1$. 

Electrons and positrons are injected at the boundary $x=0$ with the number densities $n_e = n_p=n_0$. Their velocities are initialized with a nonrelativistic Maxwellian velocity distribution with a zero mean speed and temperature $T_{cloud}=$ 100~keV and added relativistically to the mean speed $\mathbf{v}_{cloud} = (0.6c,0,0)$ of the pair cloud. The nonrelativistic thermal speed $v_{th,c}={(k_BT_{cloud}/m_e)}^{1/2}$ of the pair cloud is 0.44c or $v_{th,c}/|\mathbf{v}_{cloud}|=0.74$. A nonrelativistic Maxwellian with the temperature 100 keV does not constitute an equilibrium distribution. However, its difference from a relativistic one is small and the distribution will rapidly change in response to particle interactions with the piston and because faster pairs at the front of the pair cloud outrun the slower ones, which deforms the distribution function. Initially, most electrons and positrons move in the direction of increasing $x$. A small fraction of pairs has a negative speed in the box frame. They cross the boundary and move to negative $x$.

Electrostatic instabilities between inflowing and reflected pairs (See Fig. \ref{figure2}) are suppressed by $v_{th,c}\approx |\mathbf{v}_{cloud}|$. Weibel-type or Alfv\'enic instabilities\cite{DieckmannPPCF2019} cannot develop in an unmagnetized pair plasma if it is hotter along the simulation direction than perpendicular to it. Excluding instabilities implies that the inflowing and reflected pairs hardly interact. The total pressure, which is the sum of the thermal and ram pressures, excerted by the pair cloud on the piston thus remains constant. Absent instabilities also imply that the reflected pairs eventually cross the boundary and form a second piston at negative $x$. The energy the reflected pairs lost to the first piston implies that their pressure is reduced compared to that of the inflowing pairs; we can study the formation and evolution of pistons for two different total pressures of the pair clouds.

Our aim is not to replicate the momentum distribution of the pair cloud in Ref.\cite{DieckmannAA2019}, which also had a small velocity component of the pair cloud along the magnetic field. We want to study properties of the piston in a simplified and inexpensive setup. The pressure and particle momentum spread of the pair cloud and the thermal and magnetic pressures of the ambient plasma are nevertheless comparable to those behind the piston in Ref.\cite{DieckmannAA2019}.

We resolve the spatial domain $-150 \le x \le 150$ by 3000 grid cells and resolve the simulation time $t_{sim1}=2200$ by 35400 time steps. Protons and ambient electrons are represented by 5000 CPs per cell. We inject 5000 computational electrons and 5000 computational positrons at every time step. We refer to these injected particles as the cloud particles and distinguish at times between cloud electrons and ambient electrons.

\section{Simulation data}

\subsection{Early time evolution}
\label{subsect1}

\begin{figure*}
\includegraphics[width=\textwidth]{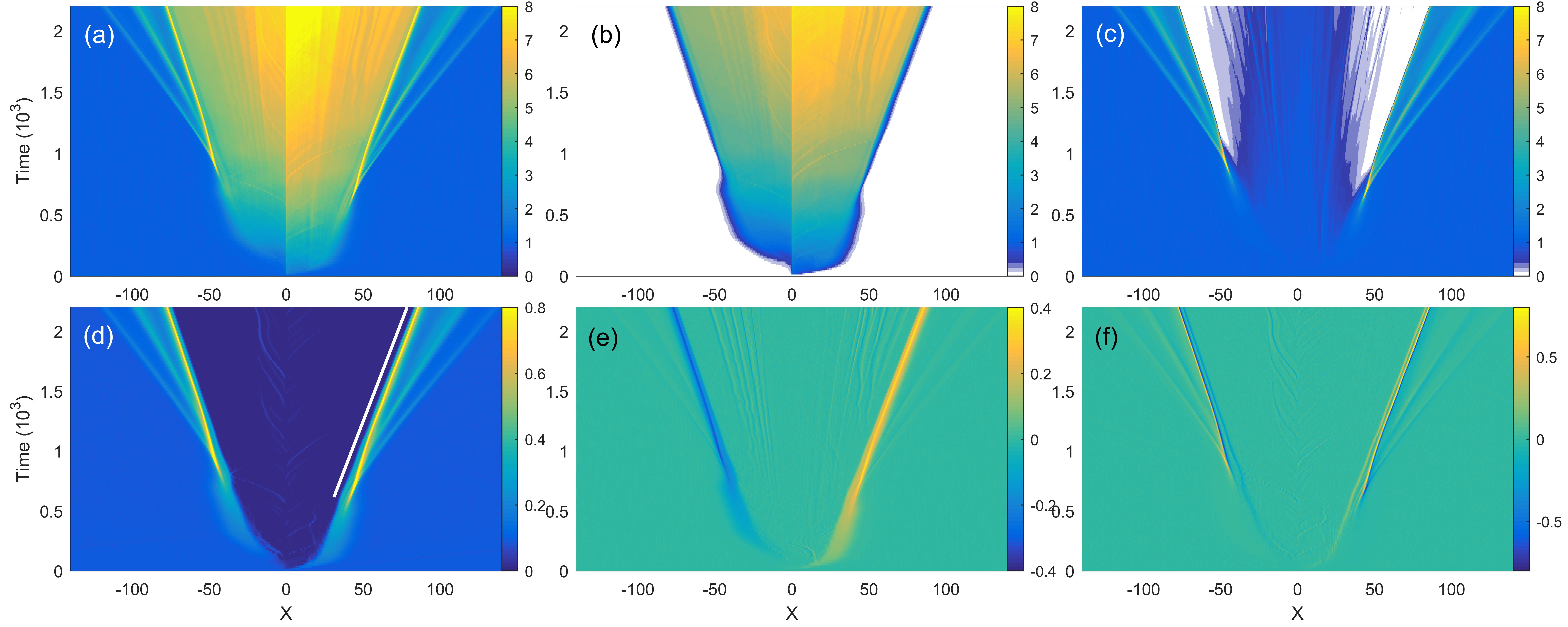}
\caption{Plasma evolution: Densities of the electrons (ambient plus cloud electrons), positrons and protons are shown in panels (a-c), respectively. Panel (d) displays the magnetic $B_y$-component. The overplotted white line marks the speed $0.033c$. Panel (e) shows the electric field component $E_x$ while the current component $J_z$  is shown in panel (f). We do not show the weak convective electric field $E_z$ and the thermal fluctuations of $J_x$. All other field and current components remain at noise levels.}
\label{figure3}
\end{figure*}

Figure \ref{figure3} displays the relevant plasma- and field quantities during the times $0 \le t \le t_{sim1}$. We have exploited the periodicity of the simulation box and shifted the boundary into the center of the figures. 
Figure~\ref{figure3}(a,~b) show how the electrons and positrons of the cloud are injected at $x=0$ and move to increasing $x$. Cloud particles are reflected by the magnetic field; the relativistic gyroradius of an electron with the speed $(|\mathbf{v}_{cloud}|+v_{th,c})/(1+|\mathbf{v}_{cloud}|v_{th,c}c^{-2})\approx 0.82c$ is $\approx 16.5$, which is comparable to the distance over which the positrons in Fig. \ref{figure3}(b) are slowed down. Cloud particles cross the injection boundary after $t\approx 200$ and flow to $x<0$. A pair cloud forms that is centred around $x=0$ and has a higher thermal pressure for $x>0$. Positron densities up to 8 are observed. Such densities are comparable to those found in the inner cocoon of the jet in Ref. \cite{DieckmannAA2019} even though the cloud we inject here has a lower density. 

The protons react to the increasing thermal pressure of the cloud close to $x\approx 0$ in Fig. \ref{figure3}(c). Protons are swiped out from intervals centred around $|x|\approx 40$ at $t\approx 500$ and accumulate at $|x|\approx 50$. Their peak density increases to about $8$ at $t=800$ and $x \approx 50$ and at $t=900$ at $x \approx -50$. Broadening outward-moving proton density pulses can be seen at later times. The proton density pulses are trailed by magnetic structures in Fig. \ref{figure3}(d) with amplitudes $B_y \approx 10B_0$. They have an electrostatic component as can be seen from Fig. \ref{figure3}(e). We show below that this electrostatic field is a consequence of having carriers of positive charge with different masses. The magnetic structure to the right in Fig. \ref{figure3}(d) travels at the speed $0.033c$. A proton, which moves with such a speed relative to a magnetic field of amplitude $10B_0$, has a gyro-radius of about 70 spatial units. This gyro-radius exceeds by far the width of the magnetic structure. Protons must thus be accelerated by the electric $E_x$ component in Fig. \ref{figure3}(e). We confirm this below. Figure \ref{figure3}(f) reveals out-of-plane currents $J_z$ of significant strength at the location of the moving magnetic field structure. 

We analyse now the effects the magnetic structure has on the plasma and examine the mechanism that generates the current $J_z$, which is connected to changes in $E_x$ and $B_y$. We turn for this purpose to the phase space density distributions of the plasma species, where those of the ambient and cloud electrons are summed up. 

\begin{figure}
\includegraphics[width=\columnwidth]{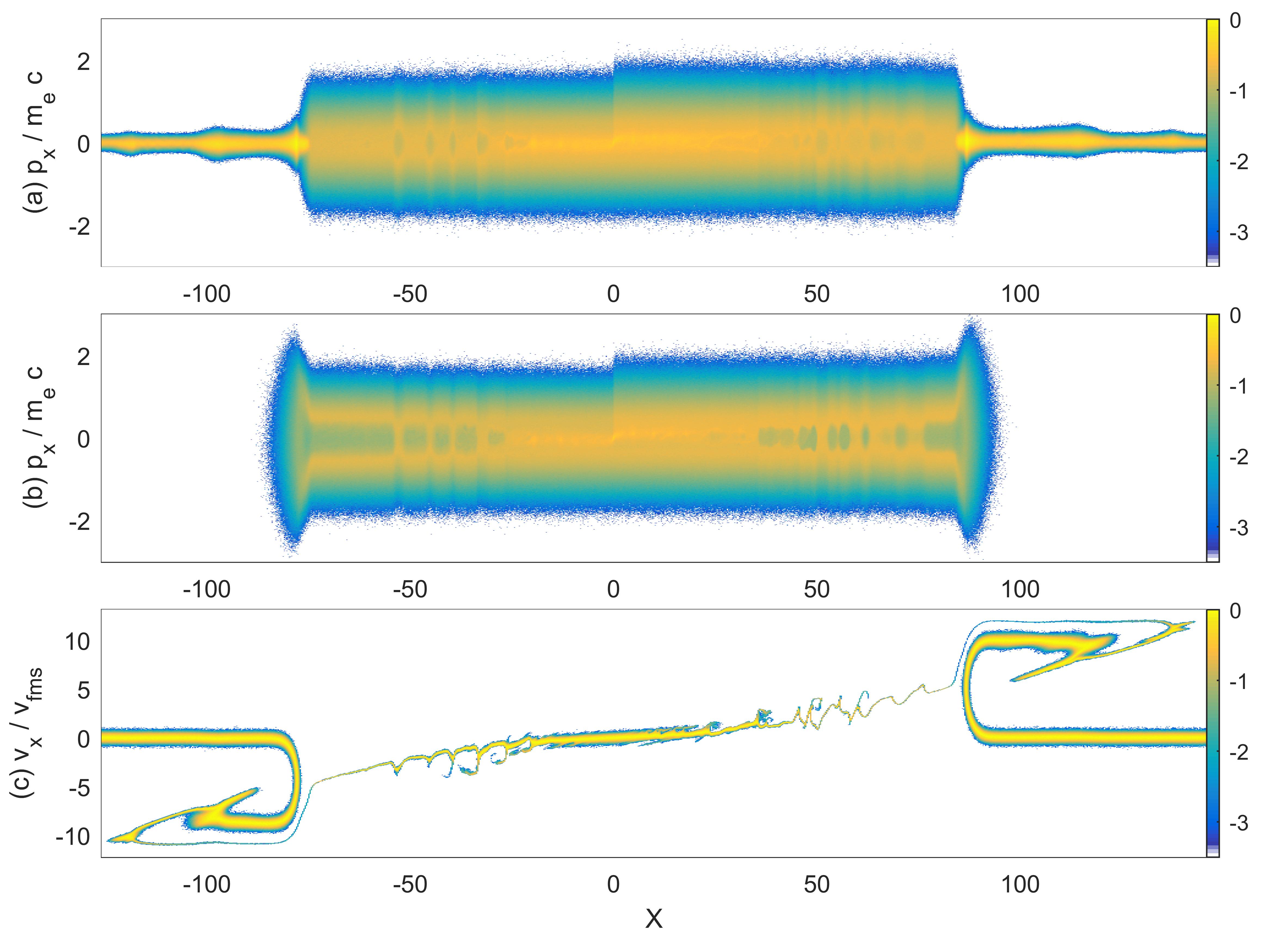}
\caption{Phase space density distributions of the electrons (a) and positrons (b) in the $x, p_x$-plane and that of the protons in the $x,v_x$-plane (c) are shown at the time $t_{sim1}$. All distributions are normalized to the peak density of the electron and proton distributions at the time $t=0$. The color scale is 10-logarithmic. Multimedia view:}
\label{figure4}
\end{figure}

Figure \ref{figure4} shows the phase space density of the electrons $f_e(x,p_x)$ and positrons $f_p(x,p_x)$ and that of the protons $f_i(x,v_x)$. The plasma evolution for all $0 \le t \le t_{sim1}$ is shown by Fig. \ref{figure4} (multimedia view). Proton velocities are normalized to the fast magnetosonic speed $v_{fms}={(c_s^2+v_A^2)}^{1/2}\approx 6 \times 10^{-3}c$ where
$v_A = B_0/(\mu_0 n_0 m_p)^{1/2}$ is the Alfv\'en speed and $c_s = {((k_B T_0(\gamma_e + \gamma_i))/m_p)}^{1/2}$ the ion acoustic speed with $\gamma_e = 5/3$ and $\gamma_i=3$. 

Figure \ref{figure4}(a) reveals a spatially almost uniform distribution in the intervals $0 \le x \le 80$ and $-70 \le x \le 0$. The electron distribution cools down at the outer boundaries of these intervals and goes over into the distribution of ambient electrons with $|x| > 100$. Positrons are heated up at $x\approx 90$ and $x\approx -80$ in Fig. \ref{figure4}(b) and they extend to larger values of $|x|$ than the hot electrons. Isocontours of the lepton distributions have an almost constant momentum for $p_x <-m_ec$ while there is a jump of the distribution at $x=0$ and $p_x \approx 2m_e c$; cloud particles have lost x-momentum after they were reflected by the boundary at $x>0$ followed by the one at $x<0$. The momentum loss is caused by the reflection of particles by an obstacle that moves in the same direction.

Figure \ref{figure4}(c) demonstrates that this momentum was transferred to the protons. Protons were not accelerated in the interval $|x| < 30$ because the magnetic structure in Fig. \ref{figure3}(d) developed outside of this interval and propagated away from it after that. Once the magnetic structure formed, it accelerated the protons at the front of the pair cloud at $x>0$ to about $11v_{fms}$ as shown by Fig.~\ref{figure4} (multimedia view). Its speed is $0.033c$ if it reflected protons specularly, which matches that of the magnetic structure in Fig. \ref{figure3}(d). Protons in the interval $x<0$ are accelerated to a lower energy, which implies that the thermal pressure of the cloud is lower at this location. The pressure drop is caused by the aforementioned momentum loss of cloud particles when they were reflected by the magnetic structure in the interval $x>0$. This momentum loss implies that the phase space densities close to the front of the electron and positron clouds cannot be symmetric about the axis $p_x=0$. Indeed, the cloud fronts must have a nonzero mean speed in order to propagate. Electromagnetic structures form for two values of the pressure that is imposed by the pair cloud on the protons. Their formation mechanism is thus robust.

\begin{figure}
\includegraphics[width=\columnwidth]{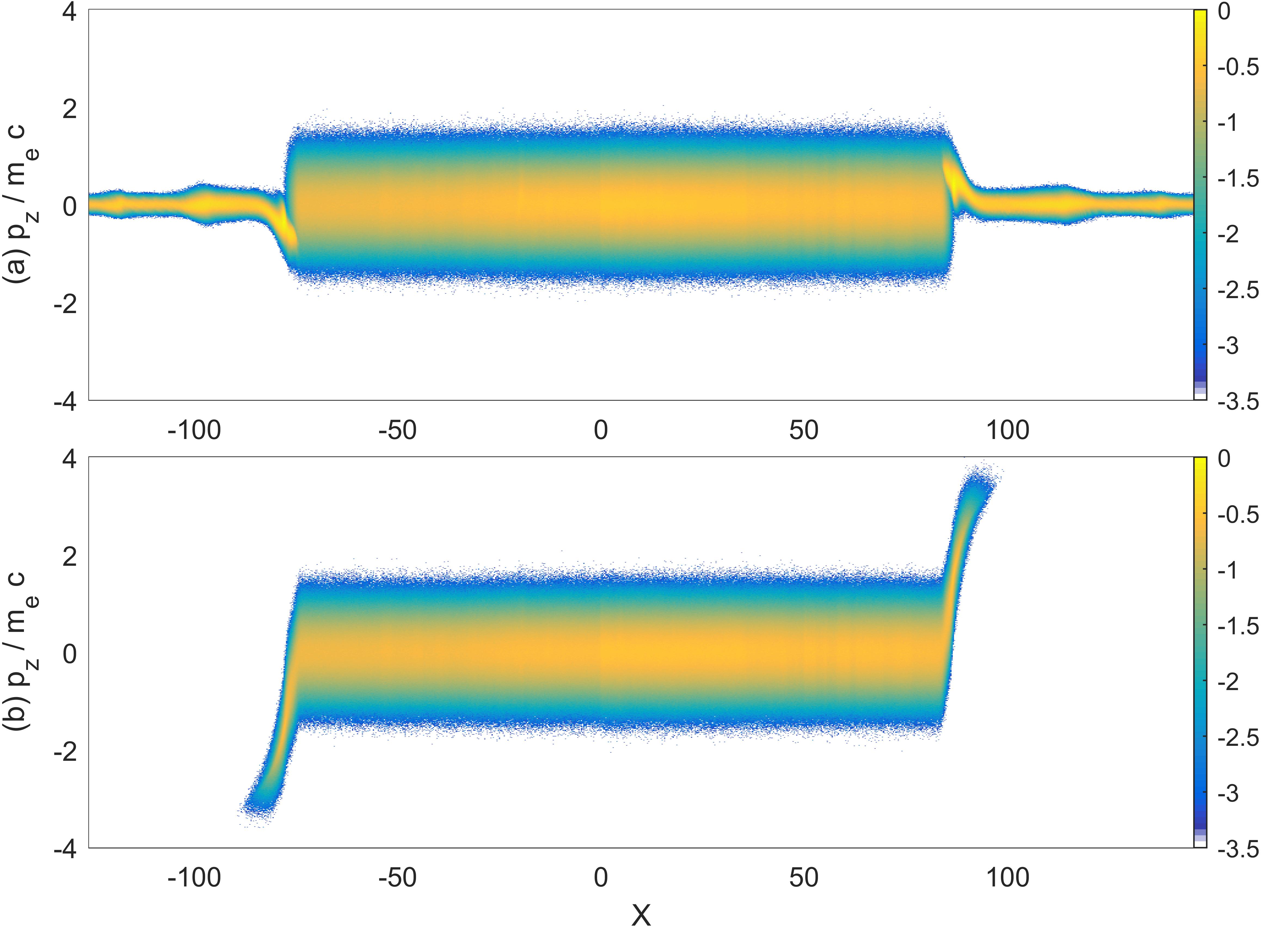}
\caption{Phase space density distributions of the electrons (a) and positrons (b) in the $x,p_z$-plane at the time $t_{sim1}$. Both distributions are normalized to the peak density of the electrons at the time $t=0$. The color scale is 10-logarithmic. Multimedia view:}
\label{figure5}
\end{figure}

Figure \ref{figure5} shows the projections onto the $x,p_z$ plane of the phase space density distributions of the electrons $f_e(x,p_z)$ and positrons $f_p(x,p_z)$ at the time $t_{sim1}$. Figure \ref{figure5} (multimedia view) animates the data for $0 \le t \le t_{sim1}$.
Both distributions show pronounced features close to the location of each magnetic structure. An energetic positron component extends to $p_z \approx 4m_ec$ at $x\approx 100$  and to $p_z \approx -3m_ec$ at $x\approx -90$. The electron distribution does not match that of the positrons; we expect a spatially varying current distribution along $z$.

\begin{figure}
\includegraphics[width=\columnwidth]{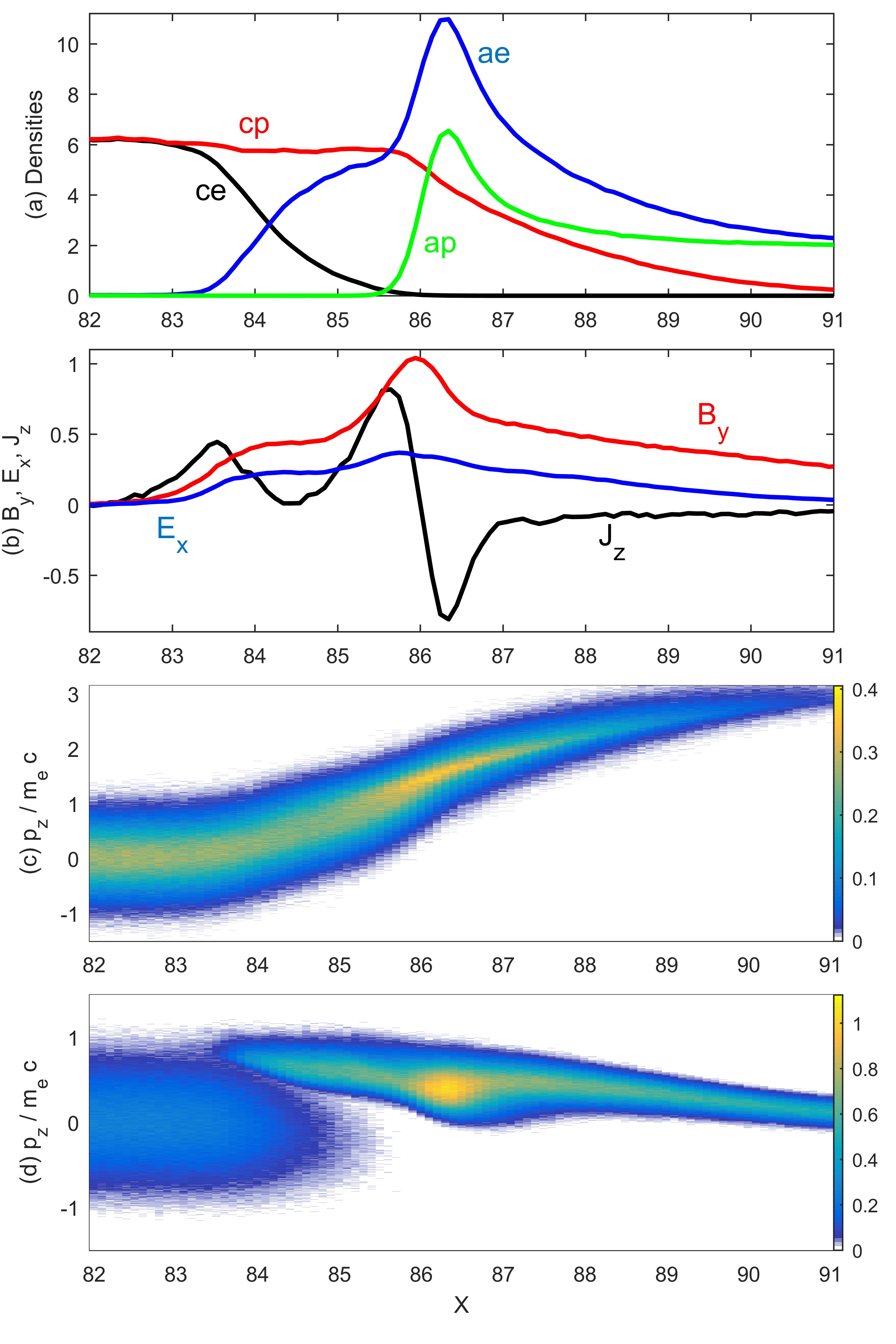}
\caption{Plasma state at the time $t=t_{sim1}$: Panel (a) shows the densities of the cloud electrons (ce, black), of the positrons (cp, red), of the ambient electrons (ae, blue) and of the protons (ap, green). Panel (b) plots $B_y$ (red), $E_x$ (blue) and $J_z$ (black). Panel (c) depicts the phase space density distribution $f_p(x,p_z)$ of the positrons and (d) that of the electrons $f_e(x,p_z)$ (ambient and cloud electrons). Both phase space densities are normalized to the same value and use a linear color scale.}
\label{figure6}
\end{figure}

Figure \ref{figure6}(a) compares the density distributions of all plasma species close to the front of the magnetic structure. We select the one that is located in the domain $x>0$ and we examine it at the time $t_{sim1}$. 
We find almost exclusively cloud particles for $x<83$ in Fig. \ref{figure6}(a). The positrons maintain their number density up to $x\approx 85.5$ while the cloud electrons are gradually replaced by ambient electrons. The density of the ambient electrons and protons increases for $x>85.5$ and both reach their peak value at $x=86.3$. About half of the positive charge density at this position is contributed by the positrons and their number density decreases to about 0 at $x=91$. 

Figure \ref{figure6}(b) shows $B_y$, $E_x$ and $J_z$ in the same interval. A small oscillation of $J_z$ is observed in the interval $82 \le x \le 84.5$ and a larger one for $84.5 \le x \le 87$. We can relate both oscillations to the distributions in Figs.~\ref{figure6}(c,~d). A rising $J_z$ at $82 \le x \le 83.5$ is tied to an increasing positive momentum of the positrons, which are accelerated by $E_x$ and deflected by $B_y$ into the z-direction. The net current decreases for $83.5 \le x \le 84.5$ due to a positive net momentum of the ambient electrons along $z$. These electrons have a small thermal gyro-radius of about 0.7 if their temperature is $T_0$ and if they rotate in a field of strength $B_0$. They undergo an $\mathbf{E} \times \mathbf{B}$-drift in the slowly changing $B_y$ and $E_x$-fields. The protons cannot undergo such a drift during the short time they interact with the magnetic structure and hence they cannot balance the current of the ambient electrons. 

The large oscillation of $J_z$ for $84.5 \le x \le 87$ is caused by the superposition of the currents arising from the spatially varying distributions of the ambient electrons and the positrons. The minimum of $J_z$ coincides with an accumulation of drifting ambient electrons (See Fig. \ref{figure6}(a)) while the positive peak at $x\approx 85.6$ is tied to comparable numbers of ambient electrons and positrons (See Fig.~\ref{figure6}(a)) and a larger drift speed of the positrons in Fig.~\ref{figure6}(c). Variations in the net current $J_z$ are responsible for the strongest changes of $B_y$ via Amp\`ere's law. 

It is not evident for now what drives the field $E_x>0$. We notice though that the electric field is monotonically rising for $82 < x < 86$ where $J_z \ge 0$ while it is decreasing in $86 < x < 91$ where $J_z < 0$. The spatial correlation of $J_z,E_x,B_y$, which we already noticed in Fig. \ref{figure3}, implies that this structure is not exclusively magnetic; we refer to it as electromagnetic piston to emphasize its resemblance to that in Ref. \cite{DieckmannAA2019}.

We gain insight into the mechanism that generates $E_x$ and ties it to $J_z$ by looking at the momentum equation of ideal magnetohydrodynamics in the comoving (Lagrangian) frame
\begin{equation}
\rho \frac{d \mathbf{v}}{dt} = \mathbf{J} \times \mathbf{B} - \nabla p_{th} = \frac{(\mathbf{B\cdot \nabla})\mathbf{B}}{\mu_0}-\nabla \left ( \frac{B^2}{2\mu_0}\right ) -\nabla p_{th},
\label{eq1}
\end{equation}
where $\rho, \mathbf{v}, p_{th}$ are the mass density, the velocity in the comoving frame and the thermal pressure of the magnetofluid. It becomes in our 1D geometry
\begin{equation}
\rho \frac{d v_x}{dt} = -\frac{d}{dx} \frac{B^2}{2\mu_0} - \frac{d}{dx}p_{th}. 
\label{eq2}
\end{equation}
If the pair cloud would expand into a vacuum that is permeated by $B_y(x) \neq 0$ then the mass density would be the sum of the mass densities of both species. No electrostatic forces would develop if both had identical distributions. The evolution of the cloud at any point $x$ would depend on how the cloud's thermal pressure gradient compares to that of the magnetic pressure. 

Figure \ref{figure4} shows that the electrons and positrons constitute a single hot fluid that works against the gradient of $B_y^2/2\mu_0$ in Fig. \ref{figure6}. According to Eqn. \ref{eq1} the pressure gradient force $\propto -B_y J_z$ points to the left and against the expanding pair cloud. However, the thermal pressure of the pair cloud exceeds the magnetic pressure, which keeps the piston moving at the speed $0.033c$. The current $J_z$, which is generated in the plasma, changes $B_y$ via Amp\`ere's law through which the electromagnetic piston moves forward. Interactions between the pair cloud and the magnetic field can thus be approximated well by the ideal MHD equations.

The piston accelerates a plasma that contains protons. It can only be this interaction that lets the ideal MHD equations break down, which yields the electrostatic field $E_x > 0$. Figure \ref{figure4}(a) suggests that the ambient electrons mix with the cloud electrons. This is however not the case according to Fig. \ref{figure6}(d) where the ambient electrons remain separated in phase space from the cloud electrons. They are pushed to increasing $x$ by the expanding cloud electrons via the magnetic pressure gradient force. Since the magnetic pressure gradient force points now to increasing $x$ we expect that the direction of $J_z$ flips, which is corroborated by Fig. \ref{figure6}(b).

\begin{figure}
\includegraphics[width=\columnwidth]{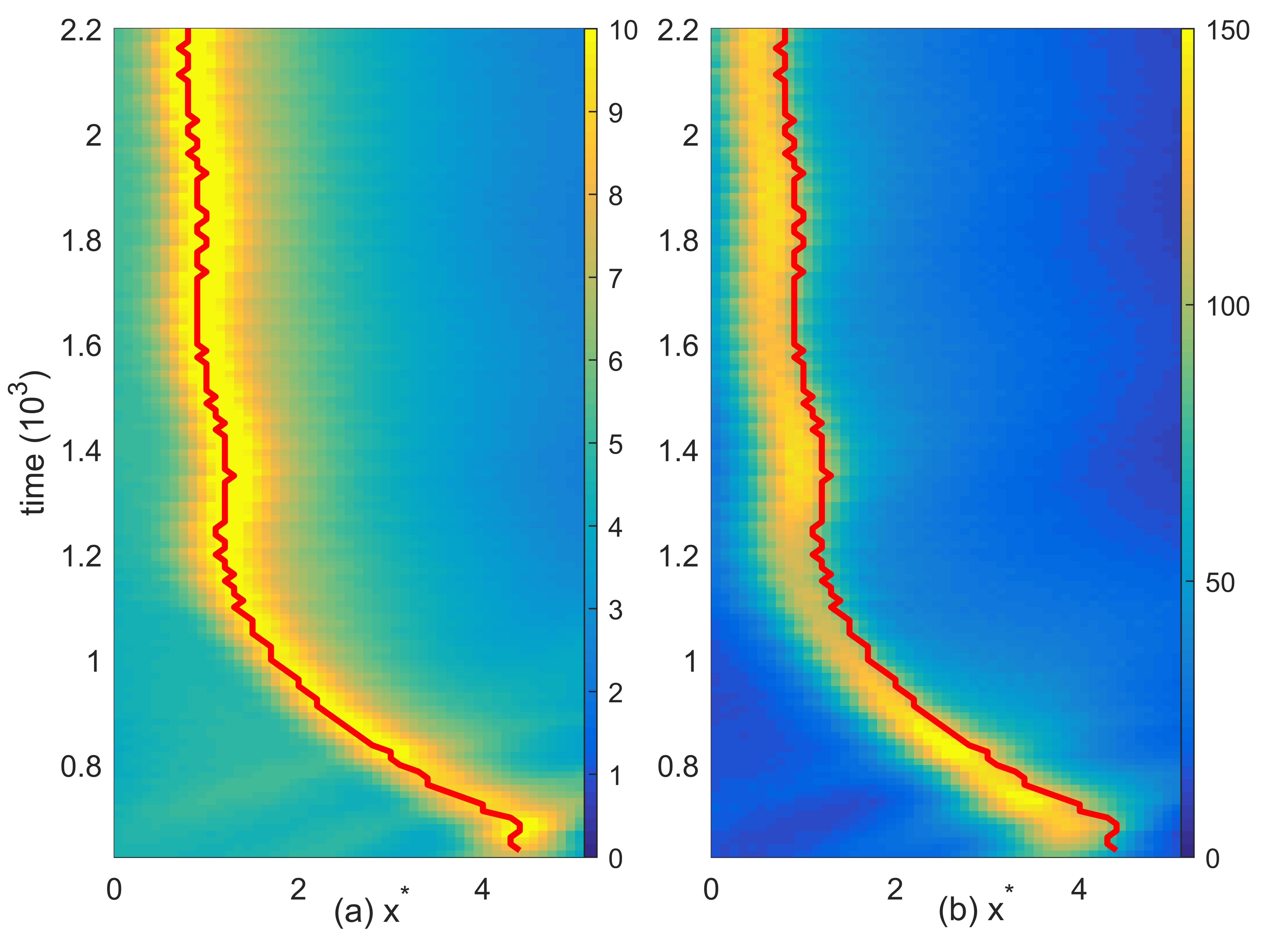}
\caption{Density of the ambient electrons (a) and magnetic pressure $P_b$ (b) in a window that moves with the speed $v_w = 0.033c$ to increasing $x$. The width of the window along $x^* = x_0 + x -v_w t$ ($x_0:$ offset along x-axis) is set such that it tracks the electromagnetic piston that moves to increasing $x$. The red curves track the maximum of the electron density in (a).}
\label{figure7}
\end{figure}

Figures \ref{figure6}(a, b) reveal that the position $x=86.3$, where the density of the ambient electrons reaches its maximum, is located in the interval where the magnetic pressure $B_y/2\mu_0$ decreases fastest. Figure \ref{figure7} shows that this is true for all times $640 \le t \le t_{sim1}$.
The current, which is associated with this motion of the ambient electrons, generates the observed electrostatic field $E_x>0$. This electric field drags positive charges with the electrons in order to maintain the quasi-neutrality of the plasma. Positrons in Fig. \ref{figure4}(b) and protons in Fig. \ref{figure4}(c) accelerate along $x$. Structures in the electron phase space density distributions in Fig. \ref{figure4}(a) (multimedia view) and Fig. \ref{figure5}(a) (multimedia view) ahead of the magnetic structure reveal that the ambient electrons are heated up while they accelerate the protons and positrons. 

The pair cloud is being kept separate from the ambient electrons and protons by the electromagnetic piston. A decreasing thermal pressure of the pair cloud coincides with an increasing pressure of the perpendicular magnetic field of the piston. Such a correlation has also been observed at the collisionless tangential discontinuity between a thermal pressure-driven blast shell of electrons and ions and a second magnetized electron-ion plasma that was initially at rest \cite{DieckmannPoP2017}. We conclude that the electromagnetic piston becomes a tangential discontinuity in the one-dimensional geometry we consider here. 

\begin{figure}
\includegraphics[width=\columnwidth]{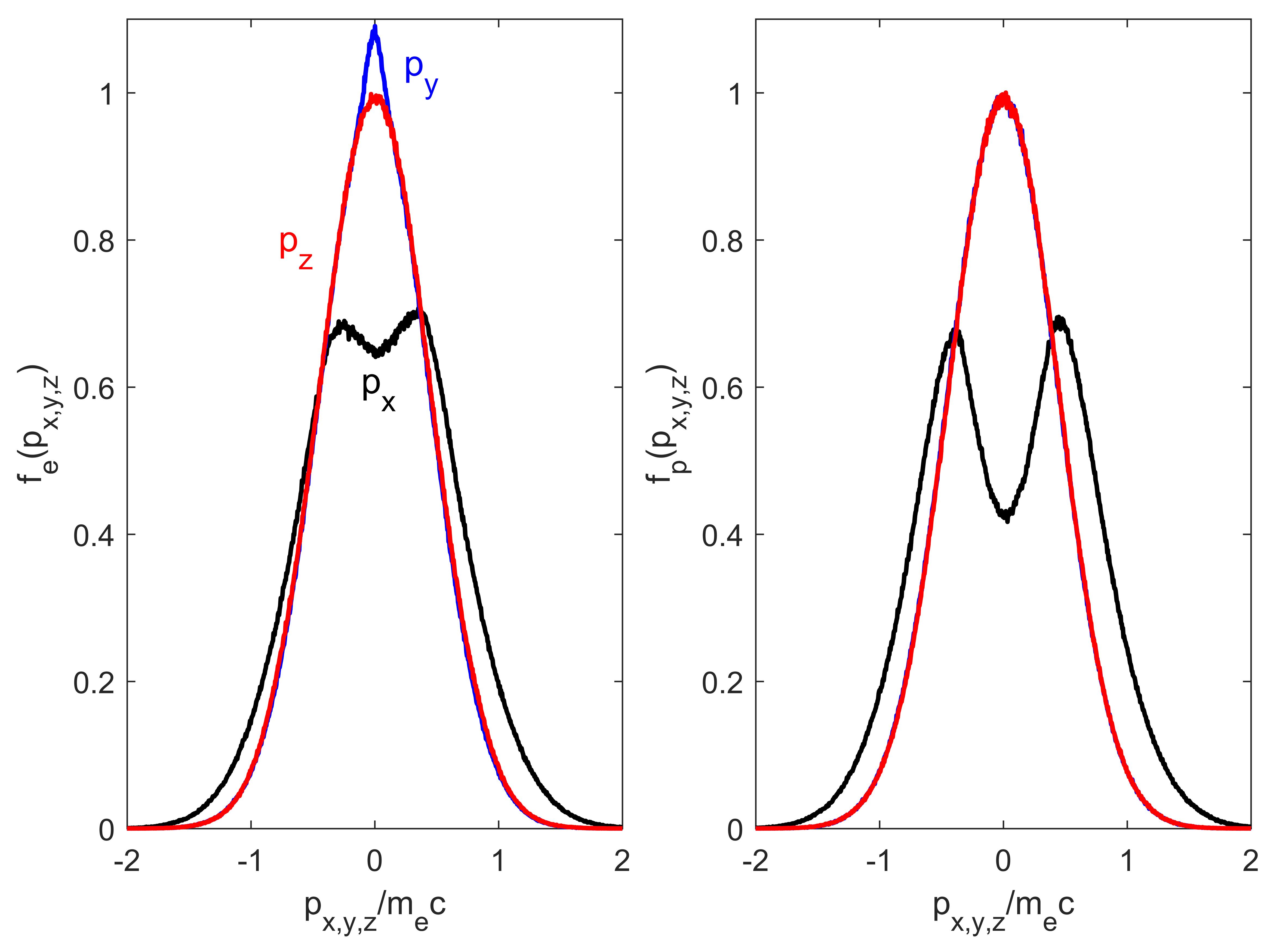}
\caption{Particle momentum distributions along $p_x$ (black curves), $p_y$ (blue curves) and $p_z$ (red curves) for electrons (a) and positrons (b). The distributions were sampled at $t_{sim1}$, they have been integrated over $50 \le x \le 82$ and normalized to the peak value in (b). The distributions $n(p_y)$ and $n(p_z)$ in (b) follow each other closely and hence we omitted plotting the blue curve. This distribution can be approximated well by a nonrelativistic Maxwellian with the temperature 100 keV.}
\label{figure8}
\end{figure}

We test if this tangential discontinuity, which confines the pair cloud in the present simulation, is also balancing the thermal pressure of the pair cloud against the sum of the magnetic pressure and the ram pressure of the protons. An estimate of the cloud temperature is needed in order to calculate its thermal pressure. We select the simulation data at the time $t_{sim1}$ and project the phase space density distributions of the electrons and positrons onto $x$ and onto the three momentum directions $p_x,p_y$ and $p_z$, respectively. The projected distributions are integrated over $50 \le x \le 82$ and shown in Fig. \ref{figure8}.

Our pair cloud has a mildly relativistic temperature and a non-relativistic Maxwellian distribution would not constitute an equilibrium distribution.
However, a fit with a Maxwellian distribution can still provide a good estimate for the cloud temperature because most particles have only mildly relativistic speeds. 

The momentum distributions of the electrons along $p_y$: $n_e (p_y)$ and along $p_z$: $n_{e}(p_z)$ in Fig. \ref{figure8}(a) and those of the positrons $n_p(p_y)$ and $n_p(p_z)$ in Fig. \ref{figure8}(b) are followed closely by a nonrelativistic Maxwellian distribution with the temperature 100 keV. This is basically the temperature the particles had when they were injected. We do not show this Maxwellian distribution because it matches $n_p(p_z)$ to within its curve thickness. A deviation of $n_e(p_y)$ from an equilibrium distribution is found at small momenta $|p_y|$. The distributions $n_{e}(p_x)$ and $n_p(p_x)$ show beams with the mean momentum $|p_x| \approx 0.75m_ec$ in a thermal background, the reason being the continuous injection of new cloud particles. These beams are more pronounced in the positron distribution than in the electron one and the energy density of the positrons is somewhat larger. Slightly different momentum distributions are not surprising because we still find some protons in the interval between the tangential discontinuities, which breaks the symmetry between the cloud electrons and positrons. Positrons close to the tangential discontinuity are also faster than the cloud electrons, which implies that the relative energy loss to the moving tangential discontinuity will be different for both species.

The cloud density is $\approx 13$ at $x=82$ in Fig.  \ref{figure6}(a). We obtain from this density and from the temperature 100 keV a thermal pressure $P_{th}$ of the cloud that exceeds $P_0$ by the factor $\approx 650$. Figure \ref{figure7}(b) shows that the magnetic pressure $P_b$ rises to about $150 P_0$ at the tangential discontinuity. The tangential discontinuity moves at a speed $v_t \approx 0.033c$, which yields a ram pressure $P_{ram} = m_pn_0v_t^2\approx 500 P_0$ that is excerted by the protons on the tangential discontinuity. A pressure balance $P_{th}=P_b+P_{ram}$ explains why the tangential discontinuity moves at an almost constant speed in Fig. \ref{figure7}.  

\begin{figure}
\includegraphics[width=\columnwidth]{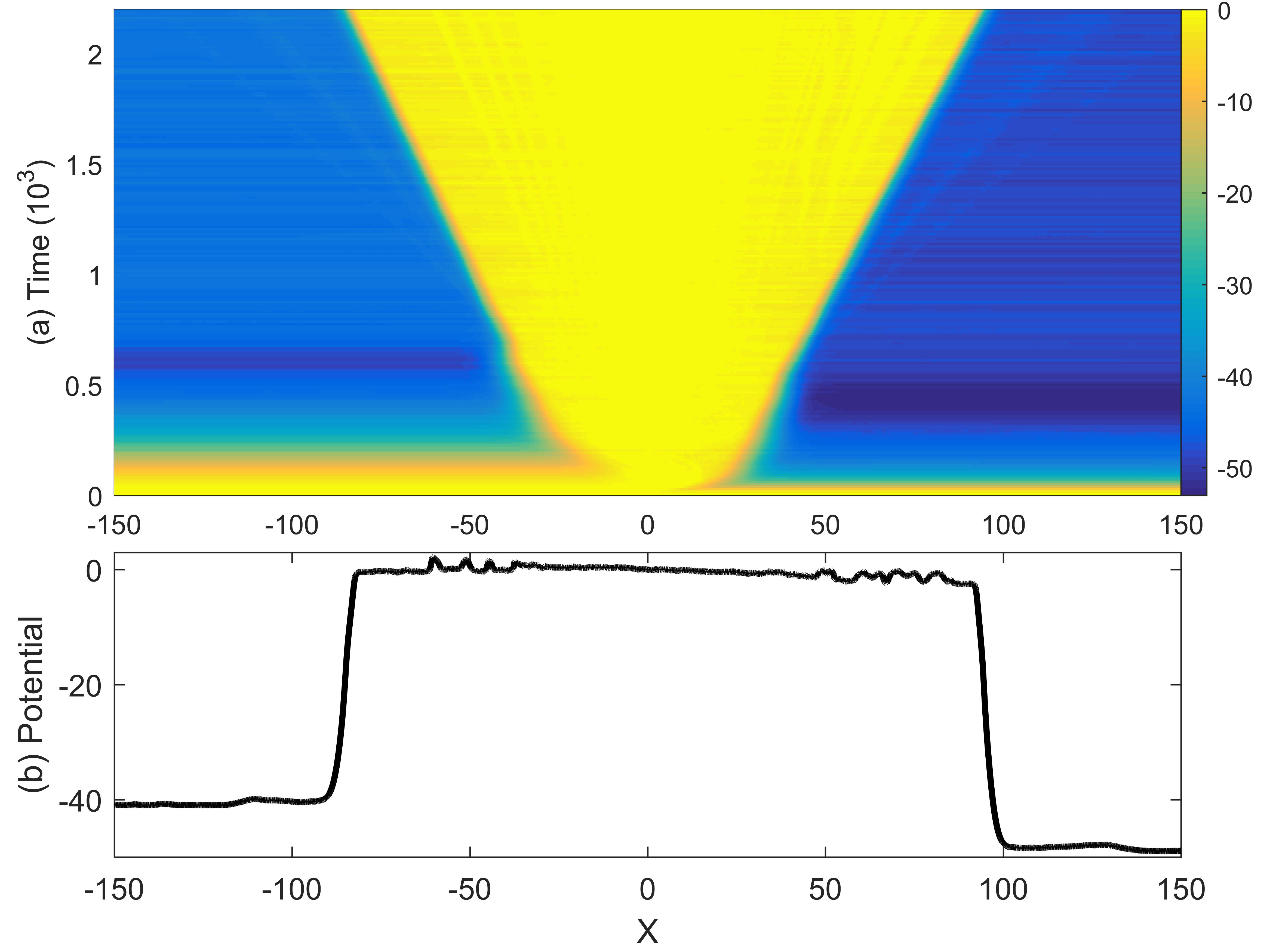}
\caption{The normalized electric field potential $\phi(x)$. Panel (a) shows it for all times. Panel (b) plots $\phi (x)$ at the time $t=t_{sim1}$.}
\label{figure9}
\end{figure}

What remains to be shown is that the electric field is strong enough to reflect the ambient protons. We neglect the changing magnetic field, which is too weak to affect the protons, and compute the electrostatic potential $E_{pot}(x_0)=-\int_{x=0}^{x_0} E_x(x)\, dx$ for all times; the reference potential is that at $x=0$. We normalize it as $\phi(x_0) = 2eE_{pot}(x_0)/(m_pv_{fms}^2)$ and drop the subscript of $x_0$. Figure \ref{figure9}(a) shows $\phi(x,t)$.

Prior to the growth of the piston's electric field (See also Fig. \ref{figure3}(e)) the potential is constant in space. A potential difference develops first in the interval $x>0$ because we inject the pair cloud into this box half. The returning pairs cross the boundary and a potential jump grows also for $x<0$. The value of $\phi (x)$ is negative outside the interval occupied by the pair cloud and hence the potential accelerates protons away from $x=0$. The potential jump is largest at $t\approx 400$ and $x>45$ and at $t\approx 600$ and $x<-50$; it overshoots its equilibrium value before the piston stabilizes. 

Figure \ref{figure9}(b) shows $\phi(x)$ at $t=t_{sim1}$. The potential jump at $x>0$ is larger and has propagated farther than its counterpart at $x<0$. We have attributed this to the larger total pressure of the pair cloud for $x>0$. The potential jump at $x\approx 100$ is about 50, which is large enough to reflect a proton moving at the speed $7v_{fms}$ relative to the piston. The potential jump at $x\approx -90$ equals 40 and can reflect protons that move at the relative speed $6.3v_{fms}$. The electric field can thus account for the reflection of the protons in Fig. \ref{figure4}(c).

It appears unphysical at first glance that the potential jumps at $x=-90$ and $x=100$ are unequal in a simulation box with periodic boundaries. However, differing potential jumps are needed because of the unequal propagation speeds of both pistons. The finite propagation speed of both pistons implies that as long as the pistons have not reached the second boundary at $x=150$ its boundary conditions do not matter for $E_x(x,t)$.  

\subsection{Late times}
\label{subsect2}

Figure \ref{figure3} has demonstrated that the piston is a stable structure on time scales of a few $10^3\omega_{pe}^{-1}$. Protons, which were accelerated by the piston, are initially too fast to interact with the ambient plasma. They will eventually be slowed down by the magnetic field of the ambient plasma with the normalized proton gyro-frequency $\omega_{ci}=eB_0/m_p\omega_{pe}\approx 4.8 \times 10^{-5}$. If we want to observe how they interact with the ambient protons, we must extend our simulation time and the box size by more than one order of magnitude. We reduce the number of particles per cell to keep the simulation time reasonable and to test if the piston is stable in a plasma with a lower statistical resolution. Our simulation box resolves the interval $-9400 \le x \le 9400$ by 187500 grid cells with the same size as in the previous simulation. Ambient electrons and protons are represented by 200 particles per cell each. We inject at $x=0$ and at each time step 200 computational particles that represent the cloud positrons and electrons, respectively. All other plasma parameters are kept unchanged. We advance the simulation until $t_{sim2}=2\pi/\omega_{ci}$, which is resolved by $2\times 10^6$ time steps.

\begin{figure*}
\includegraphics[width=\textwidth]{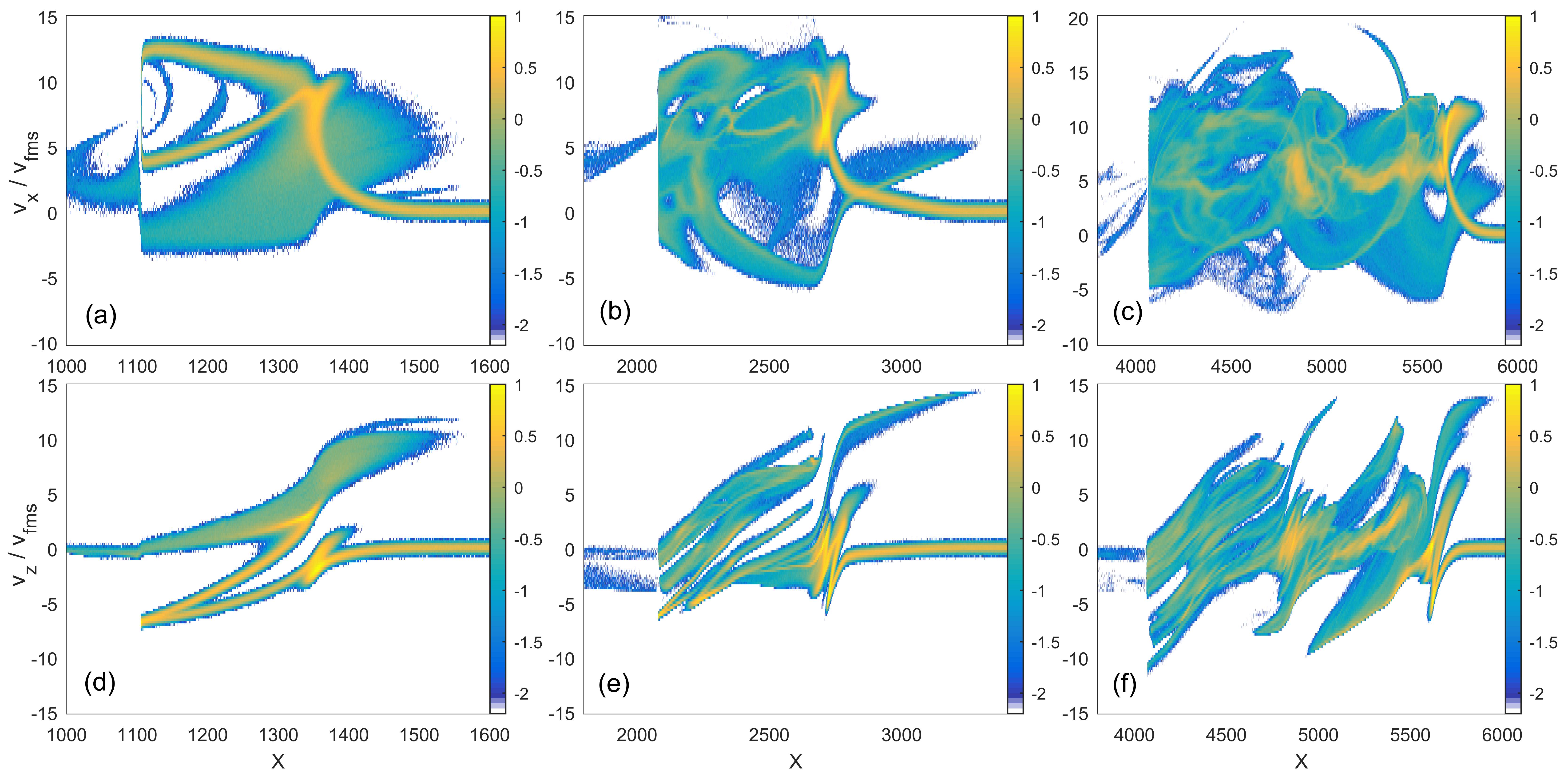}
\caption{Proton phase space density distributions at selected times. Panels (a-c) show the projections $f_i(x,v_x)$ at the times $t_{sim2}/4$, $t_{sim2}/2$ and $t_{sim2}$, respectively. Protons have performed a full rotation in the magnetic field $B_0$ at the time $t_{sim2}$. Panels (d-f) show the projections $f_i(x,v_z)$ at the same time as the panel above. All phase space densities are normalized to the peak value far upstream and displayed on a 10-logarithmic color scale.}
\label{figure10}
\end{figure*}
Figures~\ref{figure10}(a, d) show that ambient protons are confined by the piston at $x\approx 1100$. Only a few protons are observed to the left of this piston. Accelerated protons are found up to $x\approx 1550$. A dense accelerating beam of protons is observed at $x\approx 1350$. It is oriented vertically at this position and at the speed $6v_{fms}$, which evidences a magnetosonic shock. The magnetic pressure is larger behind the shock than ahead of it (not shown). It is thus mediated by the fast magnetosonic mode. Its high speed implies that it is supercritical. The presence of beams in the proton phase space densities $f_i(x,v_x)$ and $f_i(x,v_z)$ behind the shock demonstrates that the downstream protons have not yet thermalized. Protons behind the shock at $x\approx 2700$ in Figs.~\ref{figure10}(b, e) have spread over a wider phase space interval. In spite of their large thermal spread, practically all protons are confined by the piston at $x\approx 2100$. The piston has propagated until $x\approx 4200$ in Figs.~\ref{figure10}(c, f) and it confines the downstream region of the shock that is now located at $x\approx 5600$. Downstream protons cover a wide velocity interval and hardly any density accumulation is left. The piston propagates at a speed $\approx 5.4v_{fms}$ to increasing $x$ at this time.

\begin{figure}
\includegraphics[width=\columnwidth]{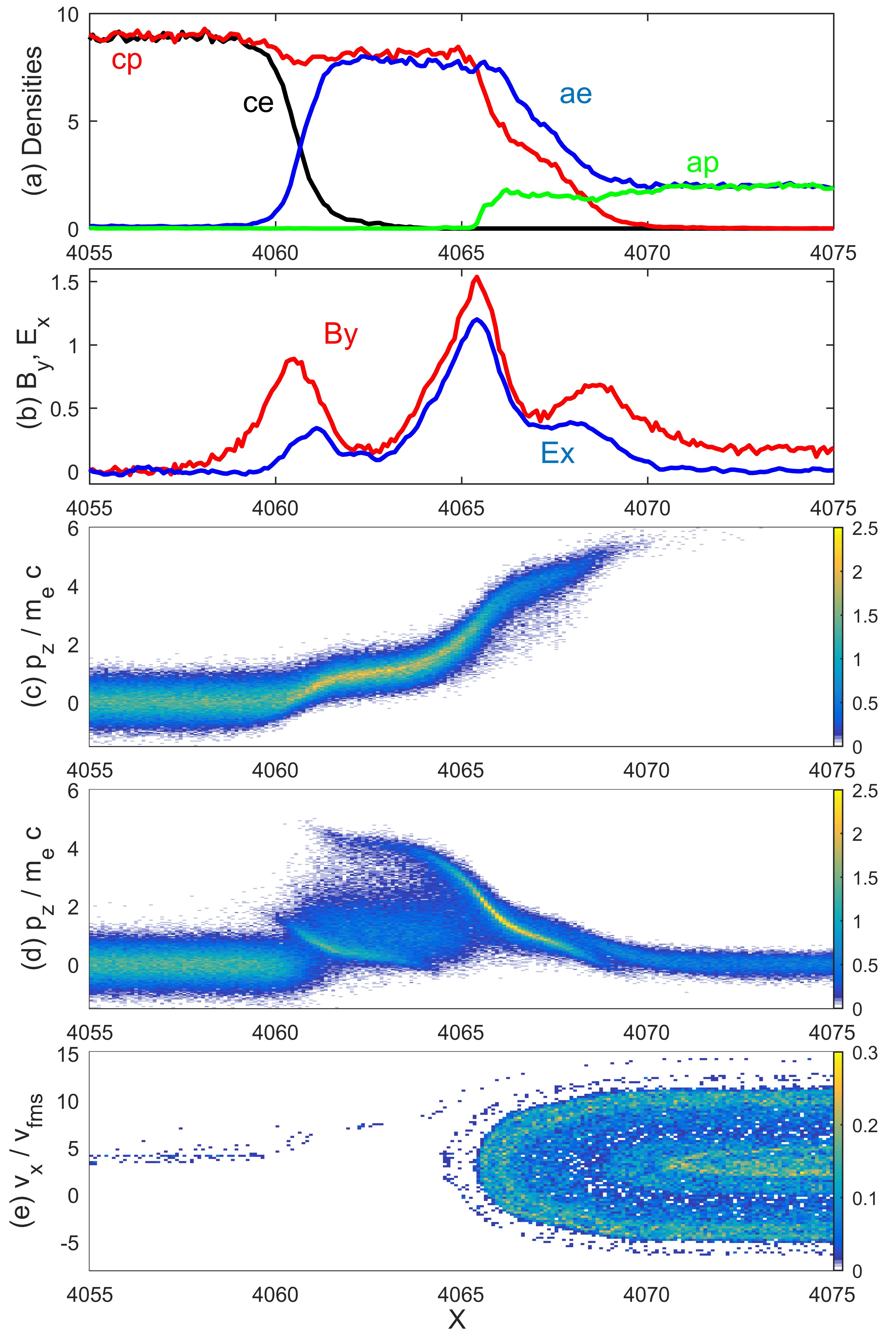}
\caption{Plasma state at the time $t_{sim2}$: Panel (a) plots the densities of the cloud positrons (cp, red) and electrons (ce, black) together with those of the ambient electrons (ae, blue) and protons (ap, green). Panel (b) plots the electric field $E_x$ (blue) and the magnetic field $B_y$ (red). The phase space density $f_p(x, p_z)$ of the positrons  is shown in (c) while (d) shows the total electron phase space density $f_e(x,p_z)$. The proton distribution $f_i (x,v_x)$ is depicted in panel (e). Phase space densities are normalized to their peak value far upstream of the shock and displayed on a 10-logarithmic color scale.}
\label{figure11}
\end{figure}

Figure \ref{figure11} presents the plasma state close to the piston at the time $t=t_{sim2}$.
The interval up to $x\approx 4058$ is occupied almost exclusively by cloud particles. Electron and positron densities are about 9. Cloud electrons are gradually replaced by ambient electrons for $4058 \le x \le 4063$ and the proton density starts to increase for $x>4065$. The density of the cloud positrons goes to zero at $x=4070$, which marks the front of the piston. Densities values $\approx 2$ of the ambient plasma ahead of the piston demonstrate that it has not yet thermalized. Density values well above 2 are expected for the downstream plasma behind a shock. The proton density rises to about 3 at larger $x$ and reaches a peak value of 16 at the shock; the large density at the shock is typical for supercritical fast magnetosonic shocks, which cannot reach a steady state in one spatial dimension. 

The amplitudes of $E_x$ and $B_y$ have grown well beyond their values in Fig. \ref{figure6}. The magnetic field is 1.5 times stronger while the amplitude of the electric field has increased by an order of magnitude. Figure~\ref{figure11}(c) shows that positrons have doubled their peak momentum along the z-direction. Their increased current leads to a stronger magnetic field $B_y$. Ambient electrons are dragged with the piston in the interval $4060 \le x \le 4070$ in Fig.~\ref{figure11}(d) and they also reach larger peak momenta. than in Fig. \ref{figure6}. New peaks in the electric field and magnetic field in Fig. \ref{figure11}(b) mark the spatial range, in which ambient electrons are trapped by the piston. The larger number of ambient electrons, which are transported by the piston along $x$, yield a larger current in this direction and, hence, a larger electric field $E_x$. Figure~\ref{figure11}(e) shows that protons are confined to the left by the largest peak of $E_x$ in Fig. \ref{figure11}(b). The piston thus still serves as a tangential discontinuity at $t=t_{sim2}$, separating the cloud particles from the ambient plasma.

\section{Discussion}

We have examined the expansion of a pair cloud in one spatial dimension. The electrons and positrons were injected at the simulation boundary $x=0$ with a mean speed 0.6$c$ and temperature 100 keV. They were injected into a plasma, which was composed of ambient electrons and protons. In contrast to a previous study with otherwise similar plasma parameters \cite{DieckmannPoP2018a}, the ambient plasma was permeated here by a spatially uniform magnetic field that was oriented orthogonally to the simulation box and the expansion direction of the pair cloud. We based our choice on the plasma conditions that were found close to the electromagnetic piston in Ref. \cite{DieckmannAA2019}.

The expanding pair cloud expelled the magnetic field and piled it up ahead of it. Eventually a tangential discontinuity formed with a magnetic field amplitude high enough to make it impossible for the ambient electrons to cross it. Ambient electrons started to drift in the magnetic field of the tangential discontinuity and were transported with it. Their net current drove an electric field that accelerated protons and positrons through which the quasi-neutrality of the plasma was maintained. 

Protons were reflected specularly by the tangential discontinuity and the energy was supplied by the inelastic reflection of the cloud particles by the moving discontinuity. The proton acceleration was much stronger than that in\cite{DieckmannPoP2018a} where protons were accelerated when ion acoustic solitary waves turned into electrostatic shocks. We observed here a speed of the reflected protons that was 10 times larger than the fast magnetosonic speed, which is comparable to that observed when the electromagnetic piston reflected protons in Ref.\cite{DieckmannAA2019}. Their interaction with the ambient protons resulted the formation of a supercritical fast magnetosonic shock on a time scale that was comparable to the inverse proton gyro-frequency. It formed far upstream of the tangential discontinuity and resembled those in Refs. \cite{SchmitzApJ2002,HoshinoApJ2002}. 

Proton reflection was the limiting factor for the propagation speed of the tangential discontinuity; it was set by the balance between the thermal pressure of the pair cloud and the sum of the magnetic pressure of the tangential discontinuity and the ram pressure the protons excerted on it. Such a balance was also observed when a blast shell of electrons and ions collided with a magnetized electron-ion plasma \cite{DieckmannPoP2017}.

In spite of its microscopic size such a tangential discontinuity is important for astrophysical outflows for three reasons. Firstly it can separate a relativistically fast outflow of electrons and positrons from an ambient plasma at rest. A flow channel devoid of ions lets the pair plasma keep its kinetic energy and its high relativistic speed for a longer time. Secondly, a boundary that separates the inner cocoon from the outer cocoon with its large mass density will slow down the lateral expansion of the jet and keep it collimated. Thirdly, the tangential discontinuity has a magnetic pressure that is comparable to the thermal pressure of the pair cloud and its magnetic fields are coherent in the plane that is orthogonal to the normal of the tangential discontinuity. Its permanent contact with the hot leptons of the inner cocoon will turn it into a source of radio synchroton emissions. 

Future work has to test the stability of the tangential discontinuity in more than one dimension and for the case that the mean velocity of the pair cloud is not aligned with the normal of the tangential discontinuity; the simulation in Ref. \cite{DieckmannAA2019} has already demonstrated a certain robustness of this structure in this case. 

\textbf{Acknowledgements} The simulation was performed on resources provided by the Swedish National Infrastructure for Computing (SNIC) through the grant SNIC2019-3-413 at the HPC2N (Ume\aa).

\end{document}